%% file: note1537.tex
\documentclass[11pt]{article}
\usepackage{graphicx}
\usepackage{epsfig}
\usepackage{rotating}
\usepackage{cite}

\newcommand{\BABARPubYear}    {06}

\newcommand{\BABARConfNumber} {29}
\newcommand{\SLACPubNumber} {11996}
\newcommand{\LANLNumber} {0000}
\def\etal {{\it et al.}}

\input babarsym

\input mysymbols

\long\def\inst#1{\par\nobreak\kern 4pt\nobreak
    {\it #1}\par\vskip 10pt plus 3pt minus 3pt}

\begin{document}

{\pagestyle{empty}

\vspace{-0.92cm}

\begin{flushright}
\babar-CONF-\BABARPubYear/\BABARConfNumber \\
SLAC-PUB-\SLACPubNumber \\
hep-ex/\LANLNumber \\
July 2006 \\
\end{flushright}

\par\vskip 5cm

\begin{center}
\Large \bf Measurement of the \bet and \betp Branching Fractions
using $\FourS\ra\BB$ Events Tagged by a Fully Reconstructed \B Meson
\end{center}
\bigskip

\begin{center}
\large The \babar\ Collaboration\\
\mbox{ }\\
\today
\end{center}
\bigskip \bigskip

\begin{center}
\large \bf Abstract
\end{center}
We report preliminary measurements of the exclusive charmless semileptonic branching fractions 
of the \bet and \betp decays. These measurements are based on $316 \invfb$ of data collected at the \FourS resonance
by the \babar\ detector. 
In events in which the decay of one \B meson to a hadronic final state is fully reconstructed, the 
semileptonic decay of the recoiling $B$ meson is identified by the detection of a charged lepton and an $\eta$ or \etap. 
We measure the branching fraction $\BR(\bet) = (0.84 \pm 0.27 \pm 0.21) 
\times 10^{-4}$, where the first error is statistical and the second one 
systematic.
We also set an upper limit on the branching fraction of $\BR(\bet) < 1.4 \times 10^{-4}$ and $\BR(\betp) < 1.3 \times 10^{-4}$ at the 90\% confidence level.

\vfill
\begin{center}

Submitted to the 33$^{\rm rd}$ International Conference on High-Energy Physics, ICHEP 06,\\
26 July---2 August 2006, Moscow, Russia.

\end{center}

\vspace{1.0cm}
\begin{center}
{\em Stanford Linear Accelerator Center, Stanford University, 
Stanford, CA 94309} \\ \vspace{0.1cm}\hrule\vspace{0.1cm}
Work supported in part by Department of Energy contract DE-AC03-76SF00515.
\end{center}

\newpage
} 

\input authors_ICHEP2006.tex

\input{intro}
\input{datasample}
\input{analysis}

\input{extraction}

\newpage
\input{systematics}
\input{conclusions}
\section{ACKNOWLEDGMENTS}
\label{sec:Acknowledgments}
\input acknowledgements.tex

\newpage

\input{references}
\end{document}

%% file: mysymbols.tex
\def\B      {\ensuremath{B}\hbox{ }}
\def\Bp      {\ensuremath{B^{+}}\hbox{ }}

\newcommand {\etap} {\etapr}
\newcommand {\az} {\ensuremath{a^0_{0}}\hbox{ }}

\newcommand {\bpiz} {\ensuremath{B^{+} \rightarrow \piz \ell^+ \nu}\xspace}
\newcommand {\bpi} {\ensuremath{B^0 \rightarrow \pim \ell^+ \nu}\xspace}
\newcommand {\bpigen} {\ensuremath{B \rightarrow \pi \ell \nu}\xspace}
\newcommand {\brhogen} {\ensuremath{B \rightarrow \rho \ell \nu}\xspace}
\newcommand {\betagen} {\ensuremath{B \rightarrow \eta \ell \nu}\xspace}
\newcommand {\betapgen} {\ensuremath{B \rightarrow \etap \ell \nu}\xspace}
\newcommand {\bomega} {\ensuremath{B^{\pm} \rightarrow \omega \ell \nu}\hbox{ }}
\newcommand {\brhop} {\ensuremath{B^0 \rightarrow \rho^+ \ell \nu}\hbox{ }}
\newcommand {\brhoz} {\ensuremath{B^{+} \rightarrow \rho^0 \ell^+ \nu}\hbox{ }}
\newcommand {\bet} {\ensuremath{B^{+} \rightarrow \eta \ell^+ \nu}\hbox{ }}
\newcommand {\betp} {\ensuremath{B^{+} \rightarrow \etapr \ell^+ \nu}\hbox{ }}

\newcommand {\bulnu}{\ensuremath{\b \rightarrow \u \ell \nu}\hbox{ }}
\newcommand {\bclnu}{\ensuremath{\b \rightarrow \c \ell \nu}\hbox{ }}

\newcommand {\Bxulnu}{\ensuremath{\B \rightarrow X_u \ell \nu}\xspace}
\newcommand {\Bxlnu}{\ensuremath{\B \rightarrow X \ell \nu}\xspace}

\newcommand {\mmiss}{\ensuremath{m_{miss}^2}\xspace}

\newcommand {\breco}{\ensuremath{B_{reco}}\xspace}

\newcommand {\bsig}{\ensuremath{B_{sig}}\xspace}

\newcommand {\D}{\ensuremath{D}\xspace}
\newcommand{\beq}{\begin{equation}}
\newcommand{\beqa}{\begin{eqnarray}}
\newcommand{\beqn}{\begin{eqnarray}}
\newcommand{\eeq}{\end{equation}}
\newcommand{\eeqa}{\end{eqnarray}}
\newcommand{\eeqn}{\end{eqnarray}}
\makeatletter
\def\slash#1{{\mathpalette\c@ncel{#1}}} 
\makeatother

\def\etal               {{\it et~al.,}}

%% file: authors_ICHEP2006.tex
\begin{center}
\small

The \babar\ Collaboration,
\bigskip

%
{B.~Aubert,}
{R.~Barate,}
{M.~Bona,}
{D.~Boutigny,}
{F.~Couderc,}
{Y.~Karyotakis,}
{J.~P.~Lees,}
{V.~Poireau,}
{V.~Tisserand,}
{A.~Zghiche}
\inst{Laboratoire de Physique des Particules, IN2P3/CNRS et Universit\'e de Savoie,
 F-74941 Annecy-Le-Vieux, France }
% ...

\end{center}\newpage

%% file: intro.tex
\section{Introduction}
\label{sec:intro}
Precise measurements of the Cabibbo-Kobayashi-Maskawa matrix \cite{CKM} 
element $V_{ub}$ can be employed to test the consistency of the 
Standard Model description of $CP$ violation.
\Vub can be extracted from exclusive charmless semileptonic \B decays 
allowing for more stringent kinematical constraints and better background 
suppression than possible with inclusive measurements. 
However, the determination of \Vub from exclusive decays is complicated by the 
presence of the strong interaction between the quarks in the initial and 
the final states. In the case of $B\to X_{u}\ell\nu$ decays, where $X_{u}$ 
is a pseudoscalar meson, and neglecting the mass of the lepton, the dynamics 
are described by a single form-factor $f(q^2)$ that depends on the square 
of the $B\to\X_{u}$ momentum transfer $q$. 
The shape of the form-factors can in principle be measured, while we have 
to rely on theoretical predictions \cite{Ball05} for their normalization.

Exclusive charmless semileptonic \B decays have been previously measured by
the CLEO \cite{CLEOpilnu}, Belle \cite{Bellepilnu} and \babar 
\cite{bad1158,piplnu,pizlnu,pilnubreco} collaborations. An extensive study 
with independent measurements of various additional charmless 
semileptonic decay modes, 
such as those involving the $\omega$, $\eta$, \etap, \az, ..., 
is important to further constrain the theoretical models and reduce the 
statistical and systematic uncertainties. 
In this paper, we present an update of our 
previous results \cite{babarichep04} on the branching fractions for the \bet 
\footnote{Charge-conjugate modes are implied throughout this paper, unless 
explicitly stated otherwise.} and \betp decay modes. 

The analysis is based on a sample of \BB events produced at the \FourS
resonance that are tagged by a fully reconstructed hadronic decay. 
The full reconstruction of the tagging \B provides a clean sample of \BB 
events and allows us to determine the flavor of the reconstructed \B meson and 
to separate \Bz and \Bp decays. 

A semileptonic decay of the recoiling \B meson is identified by the
presence of a charged lepton. 
The $\eta$ and $\etap$ mesons in the semileptonic decay are reconstructed, 
and the missing mass is calculated assuming that the $\eta$ (\etap) and the 
charged lepton are the only particles present in the recoil except for the 
undetected neutrino.
Since the momentum of the tagging \B meson is measured, a transformation into 
the rest frame of the recoiling $B$ meson can be performed. 

%% file: datasample.tex
\section{Data Sample}

The preliminary results shown here are based on a data sample 
corresponding to an integrated luminosity of 316 \invfb, containing about 
347 million \BB pairs, 
collected with the \babar\ detector \cite{babar} at the SLAC 
\pep2\ asymmetric-energy $e^+e^-$ collider \cite{pep2} operating 
at the \FourS resonance.
A Monte Carlo (MC) simulation of the \babar\ detector based on Geant4 
\cite{geant} has been used to optimize the selection criteria and to
determine the signal efficiencies and background distributions. 

%% file: analysis.tex
\section{Event Reconstruction and Selection}
\label{sec:strategy}

The analysis proceeds in three steps: first, one of the two $B$ mesons is 
fully reconstructed in hadronic decays (\breco), second, for the 
recoiling $B$ meson (\bsig) we only reconstruct a charged lepton, electron or 
muon, and then we select the exclusive decays \bet\ and \betp.
In order to minimize the systematic uncertainties due to the \breco selection 
and lepton identification, the exclusive branching fractions are measured 
relative to the inclusive semileptonic branching fraction. 

\subsection{Full Reconstruction of Hadronic $B$ Decays}
\label{sec:strategyBreco}
To reconstruct a sample of $B$ mesons, the hadronic decays $B^+
\rightarrow  \overline{D}^0 Y^{+}, \overline{D}^{*0} Y^{+}$  are selected. 
The system $Y^{+}$ consists of hadrons with a total charge of $+1$, composed 
of $n_1\pi^{\pm}\, n_2K^{\pm}\, n_3\KS\,  n_4\piz$, where $n_1 + n_2 \leq
5$,  $n_3  \leq  2$,  and  $n_4  \leq  2$. We reconstruct 
$\Dstarzb \ra \Dzb\piz, \Dzb\gamma$ and $\Dzb\ra K^+\pi^-$,
$K^+\pi^-\piz$, $K^+\pi^-\pi^-\pi^+$, $\KS\pi^+\pi^-$ and 
$\KS \ra \pi^+\pi^-$. 
The kinematical consistency of \breco\ candidates 
is checked with two variables,
the beam energy-substituted mass $\mes = \sqrt{s/4 - \vec{p}^{\,2}_B}$ and 
the energy difference $\Delta E = E_B - \sqrt{s}/2$. Here $\sqrt{s}$ is the 
total energy in the \FourS center-of-mass (CM) frame, and $\vec{p}_B$ and $E_B$
denote the momentum and energy of the \breco\ candidate in the same
frame.  
We require $|\Delta E|<3\sigma_{\Delta E}$, where $\sigma_{\Delta E}=10$ 
to $35\mev$, depending on the decay mode, is the resolution 
on $\Delta E$ for signal \breco\ events.
On average, we reconstruct one signal $\breco$ candidate in 200 \BpBm events. 

The combinatorial background from \BB\ events and  
$e^+e^-\to q\bar{q}$ ($q=u,d,s,c$) events is subtracted by performing an 
unbinned maximum likelihood fit 
to the \mes\ distribution, using the following threshold function \cite{argus}:
\begin{equation}
\label{eq:argus}
\frac{dN}{d\mes} \propto \mes \sqrt{1-x^2} \exp \left(- \xi (1 - x^2)\right),
\end{equation}

\noindent for the background (where $x = \mes/m_{\rm max}$, $m_{\rm max}$ is the endpoint 
of the curve and $\xi$ is a free parameter determined by the fit to the \mes distribution).
A Gaussian function corrected for radiation losses \cite{crystallball} peaked 
at the \B meson mass is used to describe the signal. 

\begin{figure}[!b]
 \begin{centering}
 \epsfig{file=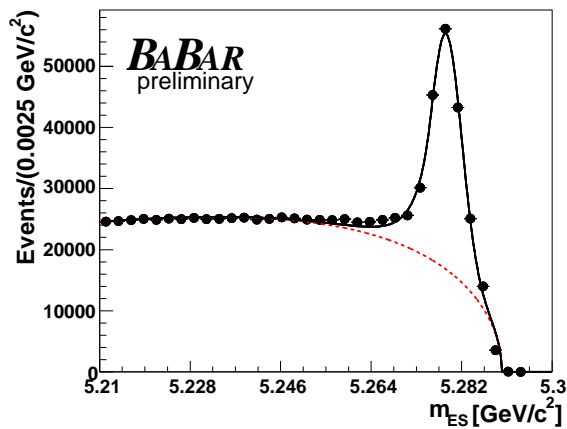,width=8cm} 
 \caption{ Fit to the \breco \mes distribution for events with a fully 
reconstructed \Bp decay, after semileptonic selection has been applied 
The fitted curve (solid line) to the data points is the 
sum of a radiation loss corrected Gaussian and a threshold function 
(dashed line) described by Eq.~\ref{eq:argus}.
\label{fig:mesfitsemilep}}
 \end{centering}
\end{figure}

\subsection{Selection of Semileptonic $B$ Decays}

The semileptonic selection identifies a charged lepton 
with a momentum $p^{*}_{\ell}$ in the \bsig rest frame
greater than $0.5\gevc$ for electrons and $0.8\gevc$ for muons. 
Electron candidates are identified using a likelihood method whose  
efficiency is about 93$\%$ and the 
hadron misidentification rate is less than 0.1$\%$. 
Muons are identified with an efficiency of about 75$\%$ and the hadron 
misidentification rate is about 3$\%$. 
We also require the lepton and the \breco\ candidate to have opposite charge 
and that the lepton track has not been used to reconstruct the \breco\ 
candidate.
Tracks are assumed to be pions if they are not identified as either 
a muon or an electron.
The number of events after the semileptonic selection is obtained with 
the \mes fit described in Section \ref{sec:strategyBreco}. 
The fit result on data is shown in Fig.~\ref{fig:mesfitsemilep}. 

\begin{figure}[!b] 
 \begin{centering} 
 \epsfig{file=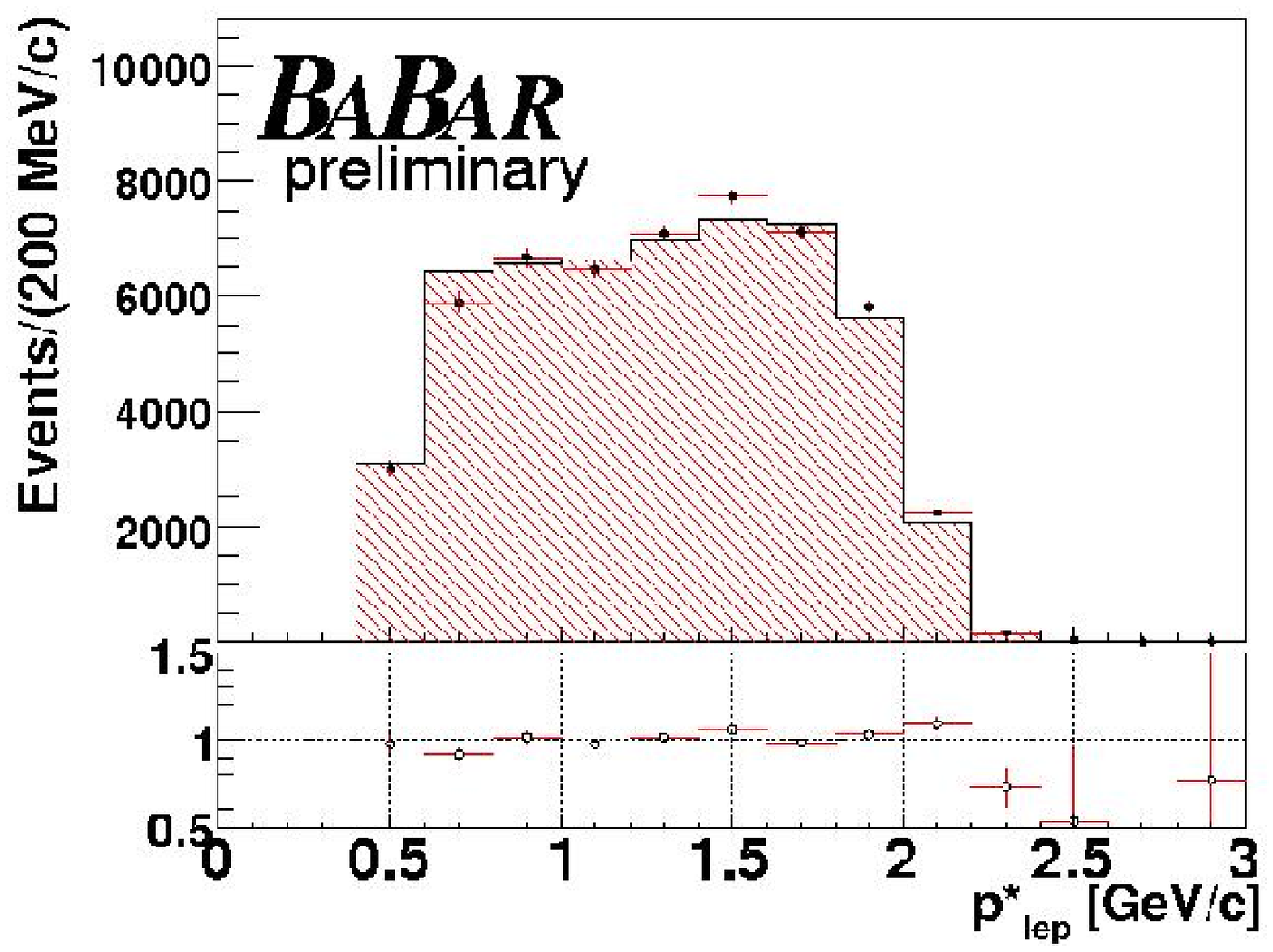,width=8cm} 
 \epsfig{file=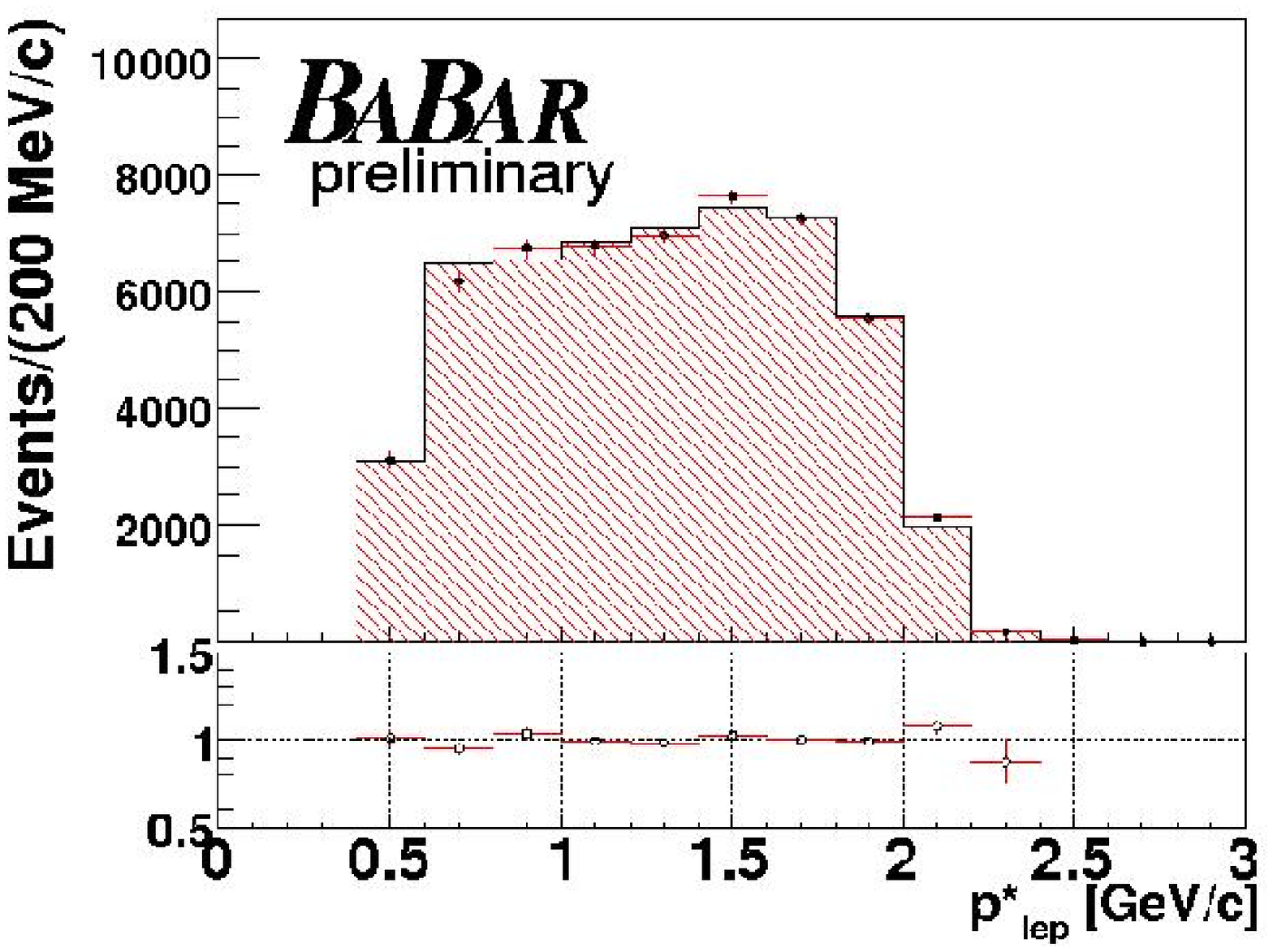,width=8cm} 
 \caption{ Distributions of the electron and muon momenta, $p^*_{\ell}$, 
computed in the rest frame of the recoiling $B$, for data points 
and Monte Carlo (histogram) for \bet (left) and \betp (right), after 
applying all cuts of the 
semileptonic selection except for the request on $p^*_{\ell}$, for events 
in which at least one $\eta$($\etap$) candidate has been found. 
The distributions are normalized to the same area. The ratio between data 
and MC simulation is shown below the histogram.  
\label{fig:pcms}} 
 \end{centering} 
\end{figure}

The distributions of the lepton momentum, $p^{*}_{\ell}$, computed in the 
recoiling \B rest frame, at this stage of the selection, are shown 
in Fig.~\ref{fig:pcms}.

\subsection{Selection of $B\to \eta \ell\nu$ and 
$B\to \eta^{'}\ell\nu$ Decays}

The \bet (\betp) decay of \bsig\ 
is reconstructed by combining $\eta$ (\etap ) candidates with the charged 
lepton.
We reconstruct $\eta$ candidates in three decay modes: 
$\eta \rightarrow \gamma \gamma$
($BF = 39.4\%$), $\eta \rightarrow \pip \pim \piz$ ($BF = 22.6\%$) and
$\eta \rightarrow \piz \piz \piz$ ($BF = 32.5\%$). The \piz candidates used 
to build the $\eta$ are defined as pairs of photons, each with an energy in 
the laboratory frame $E_{\gamma}>30\mev$, 
in the invariant mass window $110<m_{\gamma\gamma}<160\mevcc$. 
For the $\eta \rightarrow \piz \piz \piz$ channel one of 
the three reconstructed \piz mesons should satisfy additional requirements 
based on the shape of the neutral clusters of the electromagnetic 
calorimeter and a tighter cut on the invariant 
mass of the \piz ($115<m_{\gamma\gamma}<150\mevcc$). The aim of these 
additional cuts is the reduction of the combinatorial background.

We reconstruct $\etapr$ candidates in two decay modes: 
$\etapr \rightarrow \rho^{0} \gamma$ 
(\rm BF = $29.5\%$) and $\etapr \rightarrow \eta \pip \pim$ (\rm BF = 
$44.3\%$). The $\rho^{0}$ candidates used to build the \etap are reconstructed 
as pairs of charged pions with opposite charge while the $\eta$ candidates 
are selected as described above. In the $\etapr \rightarrow \rho^{0} \gamma$ 
channel we apply a cut on the momentum of the 
$\gamma$ at $p^{*}_\gamma>0.35\gevc$ to remove the background from \brhoz 
and \bclnu decays.

After these selection criteria, the dominant background is due to 
\bclnu semileptonic decays with either a real or combinatorial $\eta^{(')}$.
A good rejection variable against these events is the missing four 
momentum of the event
\begin{equation}
\label{eq:pmiss}
 p_{miss} = p_{\Upsilon(4S)} - p_{\breco} - p_{\eta(\etap )} - p_\ell, 
\end{equation}
\noindent where $p_{\Upsilon(4S)}$ is the sum of 
the four-momenta of the colliding beams, $p_{\breco}$ is the measured 
four-momentum of the \breco, $p_{\eta(\etap )}$ is the measured 
four-momentum of the $\eta$ or $\etap$ and $p_{\ell}$ is the 
measured four-momentum of the lepton. For signal events the only missing 
particle should be a single undetected neutrino, while for background 
events the missing momentum and energy in the event are due to other 
undetected or poorly measured particles. Thus, in signal events the 
resulting missing mass squared, defined as $\mmiss$ = $p_{miss}^2$, peaks 
at zero while for background events it tends to have larger values, 
and provides a 
discrimination of signal and background.

To select the decay modes of interest, the following additional selection 
criteria are applied:

\begin{itemize}

\item a cut on the invariant mass of the $\eta$ and $\etap$ candidates, different for each mode;

\item event charge balance: $Q_{tot} = Q_{\breco} + Q_{\bsig} = 0$. 
This condition rejects preferentially $b \ra c \ell \nu$ events, since their 
higher charge multiplicity implies a larger number of lost charged tracks;

\item a cut on the squared missing mass, $|\mmiss|<0.5 \gev^2/c^4$;

\item the only tracks allowed to be present in the recoil are 
the charged lepton and the tracks used to reconstruct the $\eta$ or \etap candidate;

\item for the \bet channel, the missing mass squared calculated assuming a 
\bpiz decay is required to be  $|\mmiss|_{\piz} > 1.5 \gev^2/c^4$. 
This condition rejects \bpiz events, which are the main \bulnu background 
source. 

\end{itemize}

\begin{figure}[h!]
 \begin{centering}
 \epsfig{file=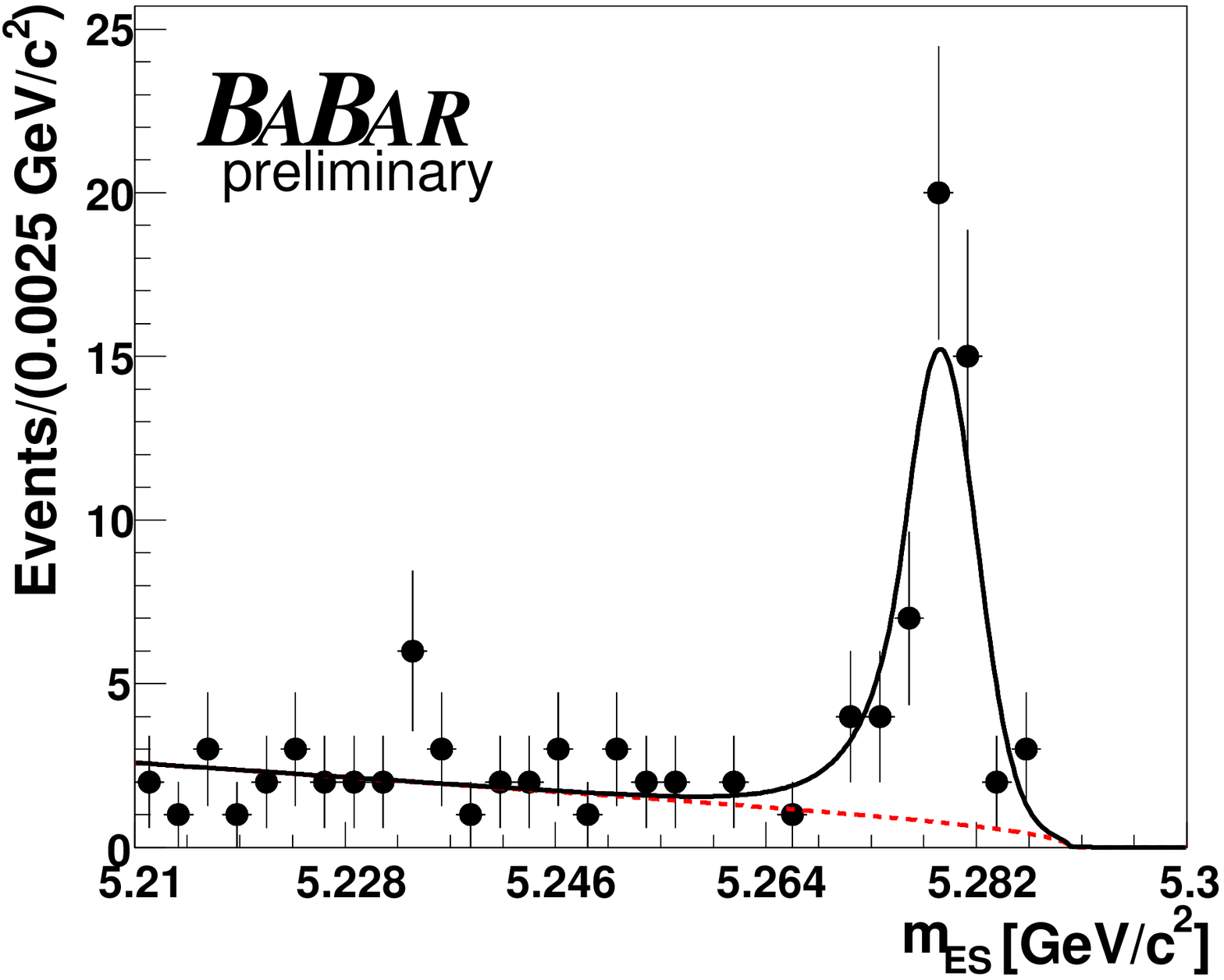,width=7cm}
 \epsfig{file=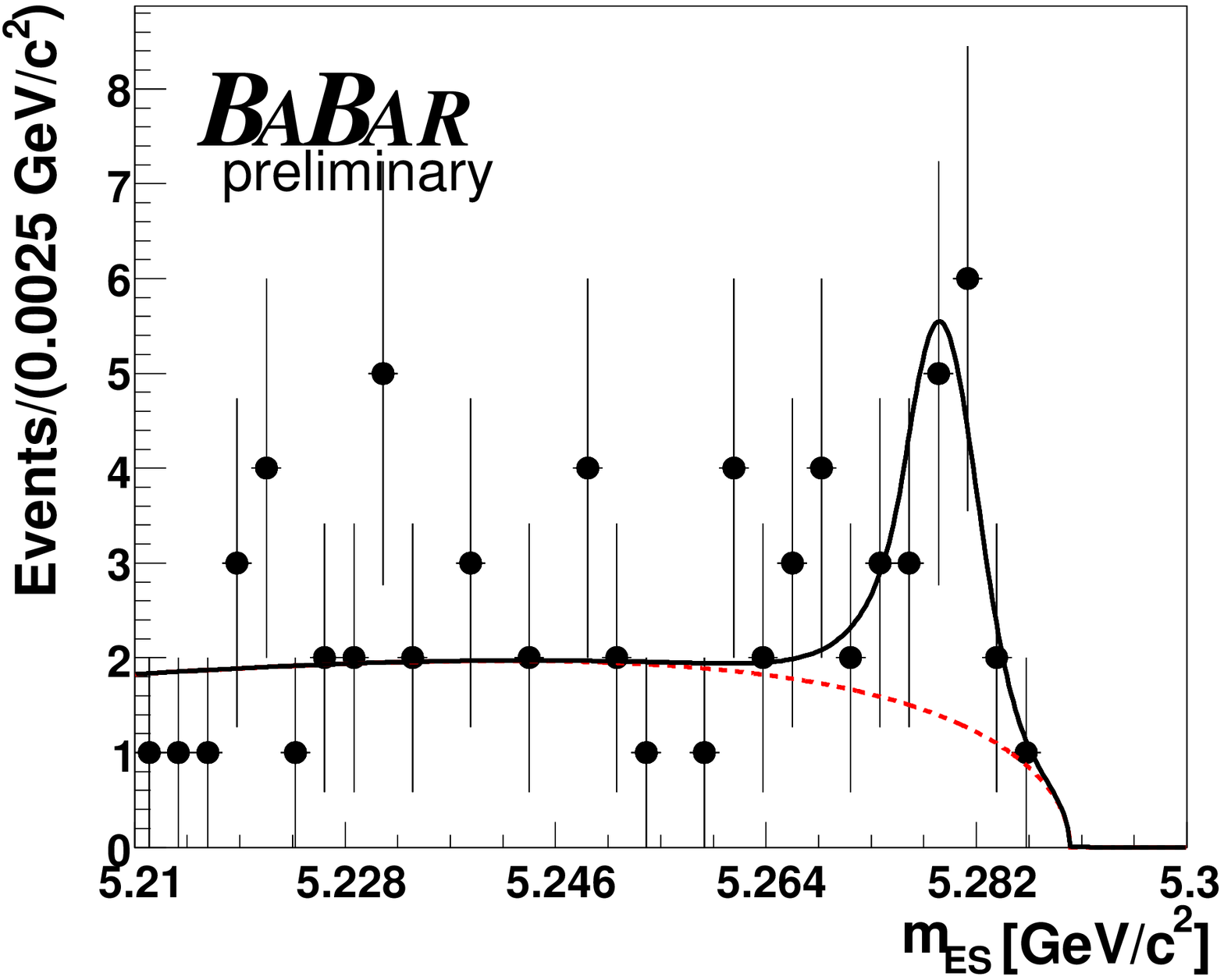,width=7cm}
 \caption{ Fit to the \mes distribution for \betagen (left) and 
\betapgen (right) candidates, after all selection criteria have been 
applied. The fitted curve (solid line) to the data points is the 
sum of a radiation loss corrected Gaussian and a threshold function 
(dashed line) described by Eq.~\ref{eq:argus}.
\label{fig:mesfitexcl}}
 \end{centering}
\end{figure}

The selection criteria described above have been optimized by minimizing the 
expected statistical error of the measurement and are 
summarized in Table ~\ref{tab:cuts}.
After all cuts have been applied we have 10-15$\%$ signal events with more 
than one $\eta$ (\etap) candidates for event. 
When several candidates remain in an event after all the cuts, the one 
with \mmiss closest to zero is chosen.
The selection efficiencies $\epsilon_{sel}^{excl}$ as estimated from the 
Monte Carlo simulation are reported in Table~\ref{tab:fitres}.
The number of events after all analysis cuts are obtained with the fit to the 
\mes distribution. The fit results on data are shown in 
Fig.~\ref{fig:mesfitexcl}.

\begin{table}[!hp]
\begin{center}
\caption{Summary of event selection for \bet and \betp. 
}
\vspace{0.1in}
\begin{tabular}{|l|l|l|} 
\hline
Selection             &  \bet & \betp \\
\hline\hline
\piz mass             & \multicolumn{2}{c|}{$110<m_{\gamma\gamma}<160 \mevcc$} \\
                      & \multicolumn{2}{c|}{$115<m_{\gamma\gamma}<150 \mevcc$ ($\eta \rightarrow \piz \piz \piz$)} \\
$\eta$ mass           &\multicolumn{2}{c|}{} \\
($\eta \rightarrow \gamma \gamma$ )                      & \multicolumn{2}{c|}{$505<m_{\eta}< 585\mevcc$ } \\
($\eta \rightarrow \pip \pim \piz$)                      & \multicolumn{2}{c|}{$530<m_{\eta}< 560\mevcc$ } \\
($\eta \rightarrow \piz \piz \piz$)                      & \multicolumn{2}{c|}{$510<m_{\eta}< 580\mevcc$ } \\ \hline

\etap mass            & & \\
($\etapr \rightarrow \rho^{0} \gamma$)                                                    &  & $930<m_{\etap}<980 \mevcc$ \\ 
($\etapr \rightarrow \eta \pip \pim$, $\eta \rightarrow \gamma \gamma$ )               &  & $940<m_{\etap}<970 \mevcc$ \\ 
($\etapr \rightarrow \eta \pip \pim$, $\eta \rightarrow \pip \pim \piz$)               &  & $935<m_{\etap}<975 \mevcc$  \\ 
($\etapr \rightarrow \eta \pip \pim$, $\eta \rightarrow \piz \piz \piz$)               &  & $910<m_{\etap}<1000\mevcc$  \\ 
$\rho^{0}$ mass             &  & $595<m_{\pip\pim}<955\mevcc$  \\ 
$\gamma$ momentum       &  & $p^{*}_\gamma>0.35\gevc$  \\ \hline
Lepton momentum       & \multicolumn{2}{c|}{$p_{el}^* > 0.5 \gevc$, $p_{\mu}^* > 0.8 \gevc$}\\
Number of leptons     & \multicolumn{2}{c|}{$N_{lepton} = 1$}\\
Charge conservation   & \multicolumn{2}{c|}{$ Q_{tot}= 0$} \\
Number of tracks      & \multicolumn{2}{c|}{no additional charged tracks} \\
Charge correlation    & \multicolumn{2}{c|}{$Q_{b(reco)} Q_{\ell} <0$} \\ 
Missing mass squared  & \multicolumn{2}{c|}{$|\mmiss|<0.5 \gev^2/c^4$} \\ \hline
\bpiz rejection       & $|\mmiss(\piz)| > 1.5 \gev^2/c^{4}$  &  \\ \hline
\end{tabular}
\label{tab:cuts}
\end{center}
\end{table}

%% file: extraction.tex
\section{Measurement of Branching Fractions}
\label{sec:BRextraction}

In order to reduce systematic uncertainties, the exclusive branching fractions 
are measured relative to the inclusive semileptonic branching fraction.

After the combinatorial background has been subtracted using the \mes fit, 
the number of inclusive \Bxlnu\ events, $N_{sl}^{meas}$, and the number of 
remaining background events, $N_{sl}^{BG}$, peaking at the \B mass in the 
\mes distribution, are related to the true number of semileptonic
decays $N_{sl}^{true}$ as:
\begin{equation}
N_{sl}^{meas} - N_{sl}^{BG} = \epsilon_l^{sl} \epsilon_t^{sl}N_{sl}^{true}.
\end{equation}
Here $\epsilon_l^{sl}$ refers to the efficiency for selecting a lepton
from a semileptonic \B decay in an
event with a hadronic \B decay, reconstructed with tag efficiency $\epsilon_t^{sl}$. 

The number of \bet (\betp) events after the \mes combinatoric background 
subtraction, $N_{excl}^{meas}$, and the number of peaking background events, 
$N_{excl}^{BG}$, 
are related to the true number of \bet (\betp) decays $N_{excl}^{true}$ as
\begin{equation} 
N_{excl}^{meas} - N_{excl}^{BG} = \epsilon_{sel}^{excl}  \epsilon_l^{excl}
 \epsilon_t^{excl} N_{excl}^{true} ,
\label{eq:NBGexcl}
\end{equation}
\noindent
where the signal efficiency $\epsilon_{sel}^{excl}$ accounts for all selection
criteria applied to the sample after the requirement of 
the presence of a charged lepton. 

The background contributions $N_{sl}^{BG}$ and $N_{excl}^{BG}$ are estimated 
using the MC simulation and scaled to the luminosity of the data sample by the 
ratio of the number of semileptonic events in data and MC. 
For $N_{excl}^{BG}$ we further rescale to match the data in a sideband 
region $1< \mmiss <4 \gev^2/c^4$. 
 
We measure the ratio of the branching fractions for \bet or \betp to the branching fraction of $\Bxlnu$ decays as
 \begin{equation}
R_{excl/sl}=
\frac{\BR(B\to\eta(\etap)\ell\nu)}{\BR(\Bxlnu)}=
 \frac{N_{excl}^{true}}{N_{sl}^{true}} = 
 \frac{(N_{excl}^{meas}- N_{excl}^{BG})/(\epsilon_{sel}^{excl})}
{(N_{sl}^{meas}- N_{sl}^{BG})} 
 \times \frac{\epsilon_l^{sl} \epsilon_t^{sl} } 
{\epsilon_l^{excl} \epsilon_t^{excl} }.
 \label{eq:ratioBR}
 \end{equation}

\noindent The ratio of efficiencies for $\Bxlnu$ and signal 
events in Eq. \ref{eq:ratioBR} is expected to be close to, but not equal to
unity.  Due to the difference in multiplicity and the different
lepton momentum spectra, we expect the tag efficiencies $\epsilon_t^{sl,excl}$
and the charged lepton efficiencies $\epsilon_l^{sl,excl}$ to be slightly different for the
two classes of events.
Using the semileptonic branching ratio $\BR(\Bxlnu) = (10.73 \pm 0.28)\%$ \cite{PDG2004}
and the ratio of the \Bz and \Bp lifetimes $\tau_{B^+}/\tau_{B^0}=1.086 \pm 0.017$ 
\cite{PDG2004}, the branching ratios $\BR(\bet)$ and $\BR(\betp)$ are derived.

\begin{figure}[h!]
 \begin{centering}
 \epsfig{file=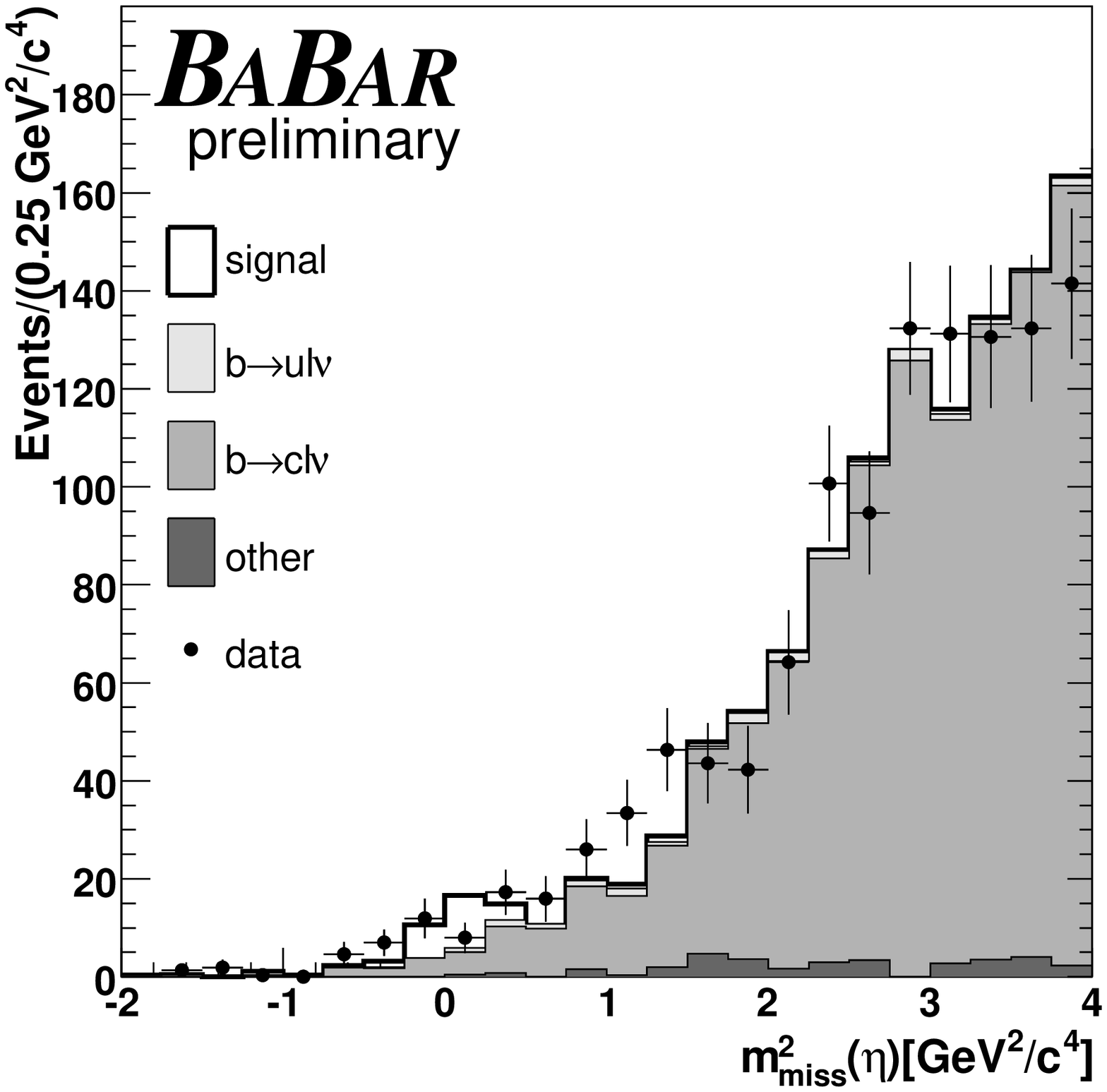,width=7.cm} 
 \epsfig{file=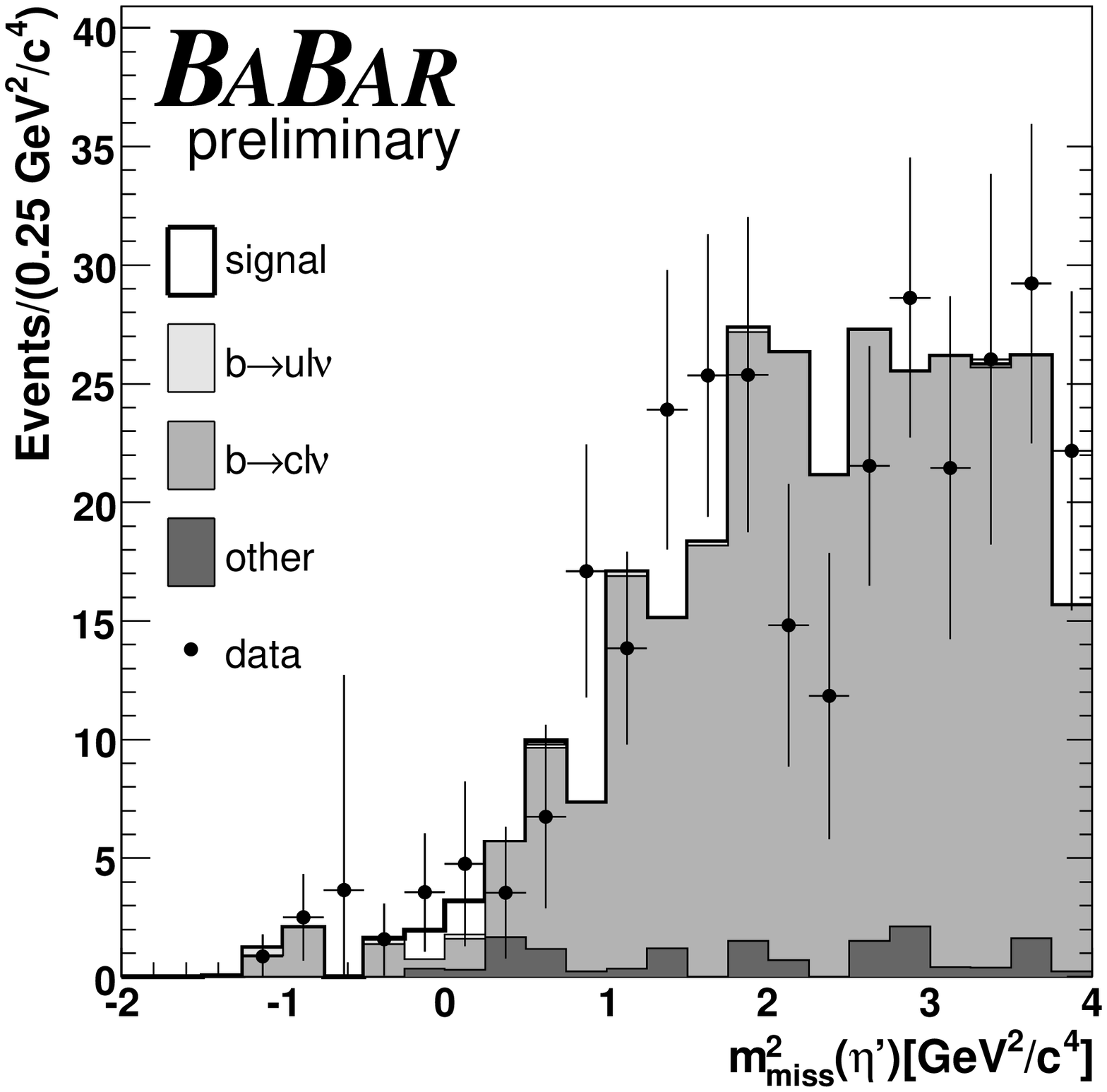,width=7.cm}         
 \caption{\mmiss distribution from data (dots) and signal and background (open and shaded histograms) contributions from Monte Carlo for \bet (left) 
and \betp (right).
\label{fig:datafit}}
 \end{centering}
\end{figure}

The results for $R_{excl/sl}$ and all the corresponding
input measurements are shown in Table~\ref{tab:fitres}.
Fig.~\ref{fig:datafit} shows the resulting data \mmiss distributions.
The signal and background components from the Monte Carlo has been scaled 
to the number of events passing the semileptonic selection and further 
rescaled by a factor 1.29 $\pm$ 0.06 for \bet and a factor 0.97 $\pm$ 0.09 for 
\betp.
We also report in Table~\ref{tab:bkgbreakdown} the contribution to the peaking 
background from the different components.

\begin{table}[h!]
\begin{center}
\caption{Measurement of  $R_{excl/sl}$ for \bet and \betp and corresponding inputs. The reported errors are statistical only.}
\vspace{0.1in}
\small
\begin{tabular}{|l|c|c|c|c|c|c|}
\hline
mode                  & $N_{excl}^{meas}$  & $N_{excl}^{BG}$  & $\epsilon_{sel}^{excl}$ & $N_{sl}^{meas}-N_{sl}^{BG}$ &$\frac{\epsilon_l^{sl} \epsilon_t^{sl} } {\epsilon_l^{excl} \epsilon_t^{excl} }$&$R_{excl/sl} [\times 10^{-3}]$  \\
\hline\hline
$\bet$               & $45.9 \pm 7.1$  & $23.8 \pm 4.9$ & $0.24 \pm 0.02$ & $109000 \pm 450$ &$ 0.88\pm 0.06 $& $0.75 \pm 0.24$ \\
$\betp$              & $14.0 \pm 5.3$ & $11.0 \pm 3.3$ & $0.10 \pm 0.01$ & $109000 \pm 450$ &$  1.05\pm 0.08   $  & $0.30 \pm 0.53$ \\
\hline
\end{tabular}
\normalsize
\label{tab:fitres}
\end{center}
\end{table}

\begin{table}[h!]
\begin{center}
\caption{Breakdown of background events for \bet and \betp. For each studied channel (columns) the background contributions from the different components (rows) are shown.}
\vspace{0.3in}
\begin{tabular}{|l|c|c|}
\hline
               & \bet & \betp \\
\hline\hline
\bpi           &  &  \\
\bpiz          & $0.77\pm0.57$ &  \\
\brhop         & $0.19\pm0.14$ &  $0.15\pm0.19$ \\
\brhoz         & &  \\
\bomega        & $0.39\pm0.29$ & \\
\bet           &  & \\
\betp          & $0.77\pm0.57$ &  \\
Other \bulnu   & $0.19\pm0.14$ & $0.45\pm0.58$ \\
\hline
$B\to D\ell\nu$ & $2.44\pm0.61$ & $1.48\pm0.51$ \\
$B\to D^{*}\ell\nu$ & $17.08\pm4.32$ & $5.94\pm2.06$ \\
Other \bclnu & $0.82\pm0.21$ & $0.89\pm0.31$ \\
\hline
 Others        & $1.21\pm1.25$ & $2.33\pm1.53$ \\
\hline
\end{tabular}
\label{tab:bkgbreakdown}
\end{center}
\end{table}  

%% file: systematics.tex
\section{Systematic Uncertainties}
\label{sec:syst}

The different sources of systematic uncertainties and their impact on the 
final results are reported in Table \ref{tab:systematics2} and briefly 
described in the following, using the same order as in the table.

The track-finding efficiency is different between data and MC simulation, 
therefore we apply a flat 0.36\% correction to the simulation in order to 
match the efficiency in data. The systematic uncertainty related to the 
reconstruction of charged tracks is determined by removing randomly a 
fraction of tracks corresponding to the uncertainty in the track finding 
efficiency.
The systematic uncertainty due to the reconstruction of neutral particles 
in the EMC is studied by varying the resolution and efficiency to match 
those found in control samples in data. Moreover we assign a systematic 
uncertainty of 1.8\% per photon and a 3.0\% uncertainty due to \piz 
reconstruction.
We estimate the systematic uncertainties due to particle identification by
varying the electron and muon identification efficiency by $\pm 2\%$ 
and $\pm3\%$, respectively, and by applying a $\pm 15\%$ uncertainty 
on mis-identification efficiency.

The uncertainty of the $B_{reco}$ background subtraction is estimated by 
using an alternative approach to evaluate $N_{sl}$, based on a binned $\chi^2$ 
fit, which is compared to the one obtained from the \mes fit to estimate the 
systematic error.
After applying the semileptonic selection, we consider the \mes distribution 
obtained from the data and from background components modeled 
with distributions taken from Monte Carlo simulation: $\Bz \bar{\Bz}$, 
$\Bp B^-$ and $e^+e^-\to q\bar{q}$ ($q=u,d,s,c$). 
We fit the background normalization on data in the \mes sideband region,
defined by $\mes < 5.26 \gevcc$.  
The relative normalization of each component is determined by a binned
$\chi^2$ fit. The $\chi^2$ function is defined as
\begin{equation}
\label{chi2}
\chi^2(C_{bkg}) = - \sum_i \left( \frac{N_i^{meas} - C_{bkg} N^{bkg,MC}_i}{\sqrt{\delta{N_i^{meas}}^2+ \delta{N_i^{MC}}^2}} \right)^2 
\end{equation}
where $N_i^{meas}$ is the number of observed events in the $i$-th bin,
$N^{bkg,MC}_i$ is the total background component, $C_{bkg}$ is the
normalization of the background component and
$\delta{N_i^{meas}}$ and $\delta {N_i^{MC}}$ are the statistical 
uncertainties for data and Monte Carlo respectively.
The normalization for $e^+e^-\to q\bar{q}$ ($q=u,d,s,c$) is fixed and the 
scaling factor is obtained from a comparison with off-peak data.
Instead, the $B^+ B^-$ and $B^0 \bar{B^0}$ components and the normalization 
of the background component are left to vary in the fit.
The total background contribution is then subtracted from the events in the
\mes signal region ($\mes >5.27 \gevcc$) in order to extract the number of
semileptonic events, separately for \Bz and \Bp and the difference is taken 
as a systematic error.

We evaluate the effect of cross-feed between \Bz\ and \Bp\ decays by 
repeating the analysis with only the $B^+B^-$ Monte Carlo sample.  
The impact of the charm semileptonic branching fractions has been estimated
by varying each of the exclusive  branching fractions within one standard 
deviation of the current world average~\cite{PDG2004}.
The effects due to exclusive \Bxulnu\ decays have been evaluated
by varying their branching fractions by $15\%$ for $\bpigen$, $30\%$ for
$\brhogen$ and by $100\%$ for the remaining decay modes.

The use of different theoretical form-factor calculations changes the 
shape of the lepton momentum spectrum for the signal and, as a consequence, 
affects the efficiencies $\epsilon_l^{excl}$, $\epsilon_l^{sl}$ and 
$\epsilon_{sel}^{excl}$. The Monte Carlo samples used in this analysis were 
generated using the ISGW2 model \cite{isgwtwo}. 
We reweight the event distributions according to the recent calculations 
by Ball and Zwicky \cite{Ball05} based on 
light-cone sum rules since, among the calculations currently available, 
these calculations result in distributions that differ most from those 
predicted by ISGW2. 
We assign the variations with respect to ISGW2 as systematic uncertainties. 
This contribution is small because the selection efficiencies for \bet and 
\betp are largely flat over the phase space.

We take into account the possible effects of the excess around 1.5 $\gevcc$ 
in the \bet case (left plot of Fig.~\ref{fig:datafit}) on the yield 
extraction. 
We varied the sideband region definition used to normalize the background 
from $1 < \mmiss < 4 \gev^2/c^4$ to $1< \mmiss < 2.5 \gev^2/c^4$, that 
corresponds to a variation on the number of background 
events $N_{excl}^{BG}$ of 11$\%$.
The difference in the branching fraction has been taken as systematic 
uncertainty.

The statistical uncertainty on the ratio of efficiencies for 
$\Bxlnu$ and signal events in Eq. \ref{eq:ratioBR}, 
due to limited Monte Carlo statistics, has been taken as a systematic 
uncertainty.    

The total systematic uncertainties on the \bet and \betp branching ratios are 
given by the sum in quadrature of all the individual contributions to the 
systematic uncertainties (Table~\ref{tab:systematics2}). 

\begin{table}[htbp]
\caption{Systematic uncertainties in the measurement of
$R_{excl/sl}$.}
\begin{center}
\begin{tabular}{|l|c|c|}
\hline
&\multicolumn{2}{c|}{Uncertainty on $R_{excl/sl}$} \\\cline{2-3}
                                   & \bet & \betp \\
\hline
\hline
Statistical error                  & $\pm 0.24$  & $\pm 0.53$  \\
\hline
\hline
Track reconstruction efficiency                & $\pm 0.04$ & $\pm 0.02$ \\
Photon resolution, \piz reconstruction        & $\pm 0.03$ & $\pm 0.03$ \\
Electron identification                        & $\pm 0.03$ & $\pm 0.01$ \\
Muon identification                            & $\pm 0.03$ & $\pm 0.02$ \\
\mes fit                           & $\pm 0.09$ & $\pm 0.04$ \\
Cross-feed $\Bz \leftrightarrow \Bp$& $\pm 0.01$ & $\pm 0.09$ \\
\hline
$B\to Dl\nu X$ and $D$ branching fractions        & $\pm 0.04$ & $\pm 0.12$ \\
\Bxulnu branching fractions                    & $\pm 0.02$ & $\pm 0.05$ \\
\hline
Form-factor~model        & $\pm 0.03$  & $\pm 0.02$ \\
\hline
Background normalization        & $\pm 0.08$ & $\pm 0.07$ \\ 
\hline
MC statistics                       & $\pm 0.12$ & $\pm 0.20$ \\
\hline
\hline
Total systematic error                      & $\pm 0.19$  & $\pm 0.27$ \\
\hline
\end{tabular}
\end{center}
\label{tab:systematics2}
\end{table}

%% file: conclusions.tex
\section{Conclusions}

We measured the branching fractions relative to the inclusive charmless semileptonic branching fraction for \bet and \betp. We obtain:

\[ 
\frac{\BR(\bet)}{\BR(\Bxlnu)} = (0.75 \pm 0.24_{stat} \pm 0.19_{syst})\times 10^{-3},
\] 
\[ 
\frac{\BR(\betp)}{\BR(\Bxlnu)} = (0.30 \pm 0.53_{stat} \pm 0.27_{syst})\times 10^{-3}. 
\] 

\noindent
Using the inclusive semileptonic branching fraction $\BR(\Bxlnu) = (10.73 \pm 0.28)\%$ and the ratio of the \Bz and \Bp lifetimes $\tau_{B^+}/\tau_{B^0}=1.086 \pm 0.017$ \cite{PDG2004}, we derive the branching fractions for \bet and \betp. We obtain:

\[
\BR(\bet) = (0.84 \pm 0.27_{stat} \pm 0.21_{syst})\times 10^{-4},
\]
\[
\BR(\betp) =(0.33 \pm 0.60_{stat} \pm 0.30_{syst})\times 10^{-4}.
\]

\noindent The significance and the upper limit has been calculated 
including all the systematic and statistical uncertainties on the 
background. 

\noindent The resulting significance is 2.55$\sigma$ for \bet and 0.95$\sigma$ 
for \betp.

For these branching fractions we get the following 90\% confidence level 
($C.L.$) upper limits:
\[
\BR(\bet) < 1.4 \times 10^{-4} ( 90\% C.L. )
\]
\[
\BR(\betp) < 1.3 \times 10^{-4} ( 90\% C.L. )
\]

%% file: acknowledgements.tex
We are grateful for the 
extraordinary contributions of our \pep2\ colleagues in
achieving the excellent luminosity and machine conditions
that have made this work possible.
The success of this project also relies critically on the 
expertise and dedication of the computing organizations that 
support \babar.
The collaborating institutions wish to thank 
SLAC for its support and the kind hospitality extended to them. 
This work is supported by the
US Department of Energy
and National Science Foundation, the
Natural Sciences and Engineering Research Council (Canada),
Institute of High Energy Physics (China), the
Commissariat \`a l'Energie Atomique and
Institut National de Physique Nucl\'eaire et de Physique des Particules
(France), the
Bundesministerium f\"ur Bildung und Forschung and
Deutsche Forschungsgemeinschaft
(Germany), the
Istituto Nazionale di Fisica Nucleare (Italy),
the Foundation for Fundamental Research on Matter (The Netherlands),
the Research Council of Norway, the
Ministry of Science and Technology of the Russian Federation, 
Ministerio de Educaci\'on y Ciencia (Spain), and the
Particle Physics and Astronomy Research Council (United Kingdom). 
Individuals have received support from 
the Marie-Curie IEF program (European Union) and
the A. P. Sloan Foundation.